\documentclass[aps,reprint,prl]{revtex4-1}
\usepackage{setspace}
\usepackage{amsfonts,amsmath,amssymb}
\usepackage{txfonts}
\usepackage{graphicx}
\usepackage{bm}
\usepackage{epstopdf}
\usepackage{verbatim}
\usepackage{xcolor}
\usepackage{units}
\usepackage[T1]{fontenc}
\usepackage[latin9]{inputenc}
\usepackage{amsbsy}
\usepackage{esint}
\nonstopmode

\newcommand{\mnras}{Mon. Not. R. Astron. Soc.}
\newcommand{\pop}{Phys. Plasmas}
\newcommand{\apjl}{Astrophys. J. Lett. }

\newcommand{\apjss}{Astrophys. J. Supp.}

\newcommand{\jplp}{J. Plasma Phys.}
\newcommand{\nature}{Nature}

\newcommand{\astronastro}{Astron. Astro.}

\newcommand{\jcompp}{J. Comp. Phys. }

\newcommand{\phrep}{Phys. Rep.}

\newcommand{\jgrsp}{J. Geophys. Res.: Space Phys. }
\newcommand{\jgr}{J. Geophys. Res}

\newcommand{\lrsp}{Living Rev. Solar Phys.}

\newcommand{\njp}{New J. Phys.}

\newcommand{\astross}{Astrophys. Space. Sci.}

\begin{document}
\title{Kinetic simulations of the interruption of large-amplitude shear-Alfv\'en waves in a high-$\beta$ plasma}

\author{J.~Squire}
\email{jsquire@caltech.edu}
\affiliation{Theoretical Astrophysics, 350-17, California Institute of Technology, Pasadena, CA 91125, USA}
\affiliation{Walter Burke Institute for Theoretical Physics, Pasadena, California 91125, USA}
\author{M.~W.~Kunz}
\affiliation{Department of Astrophysical Sciences, Princeton University, 4 Ivy Lane, Princeton, New Jersey 08544, USA}
\affiliation{Princeton Plasma Physics Laboratory, PO Box 451, Princeton, NJ 08543, USA}
\author{E.~Quataert}
\affiliation{Astronomy Department and Theoretical Astrophysics Center, University of California, Berkeley, CA 94720, USA}
\author{A.~A.~Schekochihin}
\affiliation{The Rudolf Peierls Centre for Theoretical Physics, University of Oxford, 1 Keble Road, Oxford, OX1 3NP, UK}
\affiliation{Merton College, Oxford OX1 4JD, UK}

%


\begin{abstract}
Using two-dimensional hybrid-kinetic simulations, we explore the nonlinear ``interruption'' of standing and traveling shear-Alfv\'en waves
in collisionless plasmas.
Interruption involves a self-generated pressure anisotropy removing the 
restoring force of a linearly polarized Alfv\'enic perturbation, and occurs for wave amplitudes
 $\delta B_{\perp}/B_{0}\gtrsim \beta^{\,-1/2}$ (where $\beta$ is the ratio
of thermal to magnetic pressure). We use highly elongated domains 
to obtain maximal scale separation between the wave and the ion gyroscale.
For standing waves above the amplitude limit, we find that the large-scale magnetic field of the wave
decays rapidly. The dynamics are strongly affected by the excitation  of  oblique firehose modes, which transition  into 
long-lived  parallel fluctuations at the ion gyroscale and cause  significant particle scattering. 
Traveling waves are damped more slowly, but are also influenced by small-scale parallel
fluctuations created by  the decay of 
firehose modes.
Our results demonstrate that collisionless plasmas cannot support linearly polarized Alfv\'en waves above 
$\delta B_{\perp}/B_{0}\sim \beta^{\,-1/2}$. They also provide a vivid illustration of two key aspects of low-collisionality plasma dynamics: (i) the importance of velocity-space instabilities in regulating
plasma dynamics at high $\beta$, and (ii) how nonlinear collisionless processes can transfer 
mechanical energy 
directly from the largest scales into  thermal energy and microscale fluctuations, without the need for 
a scale-by-scale turbulent cascade.

\end{abstract}
\maketitle

\emph{Introduction.---}Shear-Alfv\'en (SA) fluctuations are fundamental 
to magnetized plasma dynamics \cite{Alfven:1942hl,Cramer:2011uc,ogilvie_2016}. They are 
routinely observed in both  laboratory and space plasmas \cite{doi:10.1063/1.3592210,Bruno:2013hk}, and are the basis 
for modern theories of magnetohydrodynamic (MHD) turbulence \cite{Goldreich:1995hq,Boldyrev:2006ta,2017MNRAS.466.3918M}. 
They are also uniquely robust among plasma waves, with large-scale linear dynamics that are nearly unmodified across both kinetic and
fluid plasma models \cite{Cramer:2011uc}.

The purpose of this Letter is to explore, using hybrid-kinetic simulations, 
an important exception to this robustness. We focus on  linearly polarized  large-scale SA waves above the ``interruption limit'' \cite{Squire:2016ev,Squire:2016ev2},
\begin{equation}
 \frac{\delta B_{\perp}}{B_{0}}\gtrsim\beta^{\,-1/2},\label{eq: db limit}\end{equation}
 in a collisionless plasma.
Here $\beta\equiv8\pi p_{0}/B_{0}^{2}$ is the ratio of thermal to magnetic 
pressure, $B_{0}$ is a background magnetic field, and  $\delta B_{\perp}$ is an Alfv\'enically polarized
field perturbation. 
SA perturbations above the limit  \eqref{eq: db limit} rapidly
 transfer 
 their mechanical  energy from the largest scales to plasma  microscales and thermal energy, without the help of a turbulent cascade.
This paradigm is at odds with standard theories of Alfv\'enic turbulence in collisionless systems \cite{Schekochihin:2009eu},
and may be crucial for understanding turbulent energy dissipation in
astrophysical plasmas ranging from the intracluster medium (ICM) \cite{Rosin:2011er,PETERSON20061,2006A&A...453..447E,Zhuravleva:2014ex} to hot accretion flows  \cite{1998ApJ...500..978Q} and
high-$\beta$ regions of the solar wind \cite{Bale:2009de,Bruno:2013hk,2016JPlPh..82f5302C,BalePrep}.


The interruption of SA perturbations occurs due to the self-generation of \emph{pressure
anisotropy}, $\Delta p\equiv p_{\perp}-p_{\parallel}$ (where $p_{\perp}$ and $p_{\parallel}$
are the thermal pressures  perpendicular and parallel to $\bm{B}$). 
Pressure anisotropy is created whenever $B=|\bm{B}|$ changes in a weakly collisional plasma. Further,
if $\beta>1$, the anisotropic momentum stress $\nabla\cdot (\Delta p{\bm{B}}{\bm{B}}/B^{2})$ can be as important as, or 
even
dominate over, the magnetic tension $\nabla\cdot(\bm{{B}}{\bm{B}})/4\pi$.
 This suggests that collisionless dynamics can differ  from MHD predictions, even for large-scale perturbations satisfying
$\lambda \gg \rho_{i}$, $\tau \gg \Omega_{i}^{-1}$ (where $\rho_{i}$ and $\Omega_{i}$ are the ion gyroradius and gyrofrequency, respectively). 

  \begin{figure*}
\begin{center}
\includegraphics[width=1.0\textwidth]{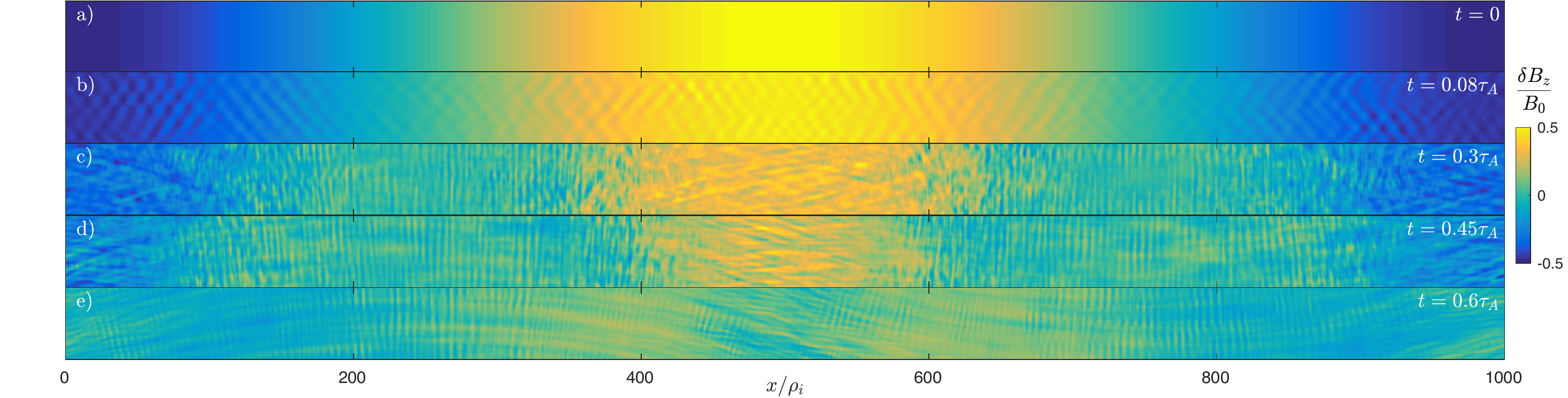}
\caption{Out-of-plane magnetic perturbation, $\delta B_{z}/B_{0}$, in a standing shear-Alfv\'en wave at $t=0$ (a), $t=0.08\tau_{A}$ (b), $t=0.3\tau_{A}$ (c), $t=0.45\tau_{A}$ (d), and $t=0.6\tau_{A}$ (e) ($\tau_{A}=10^{4}\,\Omega_{i}^{-1}$ is the linear Alfv\'en period). 
}
\label{fig: spacetime Bz standing}
\end{center}
\end{figure*}

``Interruption''
occurs when a linearly polarized SA oscillation creates an anisotropy $\Delta p = -B^{2}/4\pi$, which 
 offsets the magnetic tension and triggers the firehose instability on ion gyroscales 
\cite{1968PhFl...11.2259D,Rosenbluth:1956,Yoon:1993of,2000JGR...10510519H}. 
Even at small wave amplitudes ($\beta^{-1/2}<\delta B_{\perp}/B_{0} \ll 1$), interruption is a nonlinear effect. 
We study this behavior using hybrid kinetics (kinetic ions, fluid electrons), in three velocity  and two spatial dimensions (the latter is required to capture the 2-D oblique firehose instability, 
 which is expected to evolve in similarly three spatial dimensions). We 
consider parallel standing and traveling SA waves in the large-scale-separation limit, $\lambda_{\mathrm{mfp}}>\lambda_{A}\gg \rho_{i}$ ($\lambda_{\mathrm{mfp}}$ and $\lambda_{A}$ are the  ion mean-free path and SA wavelength), as relevant to 
many astrophysical systems (e.g., in the ICM $ \lambda_{\mathrm{mfp}}\sim 10^{11}\rho_{i}$ \cite{Rosin:2011er}).
Although the thresholds for the  oblique and parallel firehose instabilities  differ slightly 
\cite{2000JGR...10510519H,2015PhPl...22c2903K}, we  organize our discussion around the latter ($\Delta p=-B^{2}/4\pi$)  because of its 
 importance for  large-scale SA waves.


\emph{Hybrid-kinetic method.---}By treating 
 electrons as an isothermal massless  fluid, the  hybrid method removes electron kinetic
scales, plasma oscillations, and light waves from the Vlasov-Maxwell equations, reducing simulation 
cost while retaining kinetic ion dynamics \cite{1978JCoPh..27..363B,1978JCoPh..29..219H}. The equations 
consist of (i) the  Vlasov equation for the ion distribution function $f_{i}(\bm{x},\bm{v},t)$,
\begin{equation}
\frac{\partial f_{i}}{\partial t} + \bm{v}\cdot \frac{\partial f_{i} }{\partial \bm{x}}+\frac{q_{i}}{m_{i}}\left( \bm{E}+\frac{1}{c}\bm{v}\times \bm{B}\right)\cdot \frac{\partial f_{i}}{\partial \bm{v}} = 0\label{eq:Vlasov};\end{equation}
(ii) Faraday's law, ${\partial \bm{B}}/{\partial t} = -c\nabla \times \bm{E}$;
and (iii), a generalized Ohm's law,
\begin{equation}
\bm{E}+ \frac{1}{c}\bm{u}_{i}\times \bm{B}= -\frac{T_{e} \nabla n_{i}}{e n_{i}} + \frac{(\nabla\times \bm{B})\times \bm{B}}{4\pi q_{i} n_{i}}. \label{eq: Ohms}\end{equation}
Here, $q_{i}$ and $m_{i}$ are the ion's charge and mass, $\bm{E}$ is the electric  field, $c$ is the speed of light, and $T_{e}$ is the electron temperature. The ion density $n_{i}(\bm{x})\equiv \int d\bm{v}\, f_{i}$ and bulk 
velocity $\bm{u}_{i}(\bm{x})\equiv\int d\bm{v}\, \bm{v}f_{i}$ are calculated from  $f_{i}$,
 closing the system.


We use the second-order-accurate particle-in-cell  (PIC) code, \emph{Pegasus}  \citep{Kunz:2014hm}. We employ the $\delta f$ method \citep{Chen:2003de}, which
evolves $\delta f = f - f_{0}$ rather than $f$ itself, and take $f_{0}$  to be an isotropic Maxwellian. This reduces particle noise by $\sim\!(\delta f /f_{0})^{2}$, making it optimal for simulation 
 of high-$\beta$ plasmas, where  very small ($\ll 1/\beta$) deviations from a Maxwellian 
 distribution must be accurately resolved.

 \emph{Simulation set up.---}We consider two initial conditions, which vary initially only on  large scales. These are (i) a parallel standing  SA wave 
 initiated by a magnetic perturbation, and (ii) a  parallel traveling SA wave.  We focus on 
 the standing wave because of its relevance to situations where $\langle dB/dt\rangle\neq0$, e.g.,  Alfv\'enic turbulence ($\langle\cdot\rangle$ represents a spatial average). Although also important, we leave study 
 of initial Alfv\'enic velocity perturbations to future work, due to
 the  larger domains required to capture mirror instability dynamics  \cite{Melville:2015tt,Rincon:2015mi}.
The initial ion distribution function is an isotropic Maxwellian with $T_{e} = T_{i}$, with a background
 magnetic field $\bm{B}=B_{0}\hat{\bm{x}}$ and $\beta_{i}=8\pi n_{i}T_{i}/B_{0}^{2}=100$. Our domains have 
 width $L_{y}=50\rho_{i}$ and lengths up to  $L_{x} = 1000\rho_{i}$, to maximize scale separation between the SA wave 
 and  microscale dynamics.  We use a spatial resolution of $\Delta x = 0.3125 \rho_{i}$ and    $N_{\mathrm{ppc}} = 4096$ particles per cell (ppc) for the two main simulations in this Letter. 
 We initialize with $\lambda_{A}=L_{x}$ in the out-of-plane field, 
$\delta B_{z} = -\delta b\, B_{0} \cos (2\pi x/\lambda_{A} )$,
 and, for the traveling wave, a corresponding velocity perturbation, $\delta u_{z}  = \delta b\, v_{A} \cos (2\pi x/\lambda_{A} )$. In each case, we take the wave amplitude $\delta b= 0.5$, which is well above the interruption limit $\delta b_{\mathrm{max}} \approx 2 \beta^{\,-1/2}$ \cite{Squire:2016ev}. Within
 the MHD model, these initial conditions would create continuing sinusoidal SA 
 oscillations  of period $\tau_{A} = 2\pi/\omega_{A} = \sqrt{\beta_{i}}\lambda_{A}/\rho_{i} \Omega_{i}^{-1}$ (modified slightly by compressibility \citep{Squire:2016ev2}).

Due to the wide range of time and space scales involved in this problem, careful numerical 
tests are crucial. In addition to previous
\emph{Pegasus} tests \cite{Kunz:2014hm}, we  tested the numerical parameters required to accurately propagate long-wavelength linear SA waves
(with $\lambda_{A}/\rho_{i}=50$ to $1000$, $\delta b=0.05$). These tests  demonstrated   that  high  $N_{\mathrm{ppc}}\propto \lambda_{A}/\rho_{i}$ is required for large wavelengths, due to the build up of PIC noise over long simulation times. For production runs,  $N_{\mathrm{ppc}}=4096$ was 
chosen based on these requirements.
We also tested the convergence (with $N_{\mathrm{ppc}}$) of  nonlinear  standing waves  at $\lambda_{A}/\rho_{i}=250$, and their dependence on $\lambda_{A}/\rho_{i}$ over the range $\lambda_{A}/\rho_{i}=125$ to $1000$. We observed broadly similar dynamics over this range.

\begin{figure}
\begin{center}
\includegraphics[width=1.0\columnwidth]{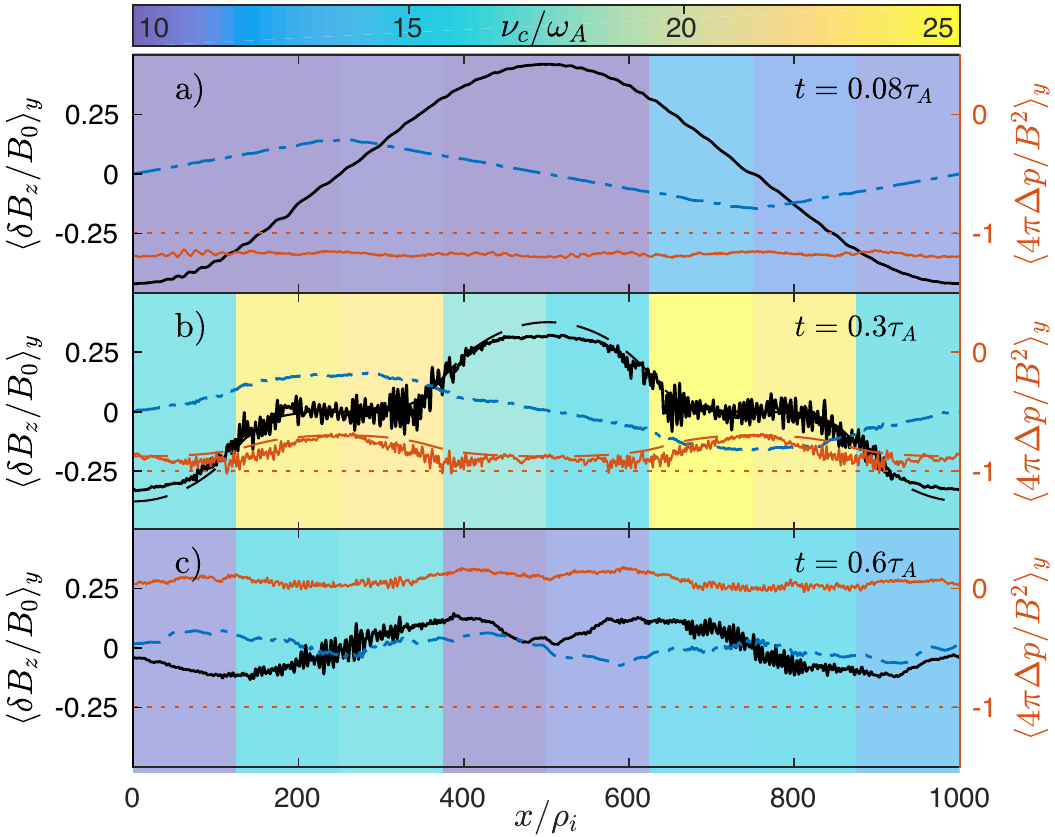}
\caption{Evolution of the standing wave from Fig.~\ref{fig: spacetime Bz standing}: we show the $y$-averaged $\delta B_{z}/B_{0}$ (black line, left axis), 
$\delta u_{z}/v_{A}$ (blue dot-dashed line, left axis), and firehose parameter $4\pi\Delta p/B^{2}$ (red, right axis; the dotted red
line shows $\Delta p=-B^{2}/4\pi$), at the  times illustrated in Figs.~\ref{fig: spacetime Bz standing}(b,c,e). The background color shows the effective  collisionality $\nu_{c}/\omega_{A}$ caused 
by particle scattering from microscale fluctuations,  measured over the time intervals $t/\tau_{A}\in[0.07,0.15]$ (a), $t/\tau_{A}\in[0.2,0.4]$  (b), and $t/\tau_{A}\in[0.55,0.65]$ (c). In (b), we also show (dashed lines)  $\delta B_{z}/B_{0}$ and $4\pi\Delta p/B^{2}$ for a decaying SA standing wave in a Braginskii model at $\beta=100$, $\nu_{c}/\omega_{A}\approx 10$ ($\delta u_{z}/v_{A}$ is omitted for clarity).
}
\label{fig: collisionality and Bz standings}
\end{center}
\end{figure}

\emph{Shear-Alfv\'en standing wave.---}Figure~\ref{fig: spacetime Bz standing}
 shows the 
spatiotemporal evolution of  $\delta B_{z}$ for a 
standing SA wave with $\lambda_{A}/\rho_{i}=1000,\:\tau_{A}\Omega_{i}=10000$ ($\omega_{A}/\Omega_{i} = 2\pi\times 10^{-4}$).  The pictured snapshots are chosen to illustrate  the distinct phases
of nonlinear wave evolution. These are: (i) initial field decrease, which creates a negative anisotropy $\Delta p<-B^{2}/4\pi$, nullifying magnetic tension and triggering the firehose instability; (ii) eruption of oblique firehose modes \cite{Yoon:1993of,2000JGR...10510519H,Kunz:2014kt} which push the wave back above
$4\pi\Delta p /B^{2} = -1$; (iii)   decay of oblique firehose modes into smaller-scale ($k_{\perp}=0$, $k_{\parallel}\rho_{i}\sim 1$) fluctuations that  scatter particles and cause the large-scale $\delta B_{z}$ to decay; and (iv),  dissolution of the wave into freely oscillating SA waves below the limit \eqref{eq: db limit},
which can oscillate freely.
Of these stages, (iii) is notably different from the predictions of 1-D Landau-fluid (LF) models \citep{Squire:2016ev,Squire:2016ev2}.

Figure~\ref{fig: collisionality and Bz standings} shows  1-D ($y$-averaged) wave profiles.
Because of heat fluxes  \cite{Squire:2016ev2}, as   $\Delta p$ decreases initially it is nearly homogenous in space [Fig.~\ref{fig: collisionality and Bz standings}(a)]. This causes oblique firehose modes to erupt suddenly across the entire wave  [Fig.~\ref{fig: spacetime Bz standing}(b)] at $t/\tau_{A}\approx0.075$. These growing modes  cause $\Delta p$  to increase  \cite{Schekochihin:2008en} back into the stable regime $\Delta p > -B^{2}/4\pi$ by $t/\tau_{A}\approx 0.085$, where it stays 
until the SA wave decays.


\begin{figure}
\begin{center}
\includegraphics[width=1.0\columnwidth]{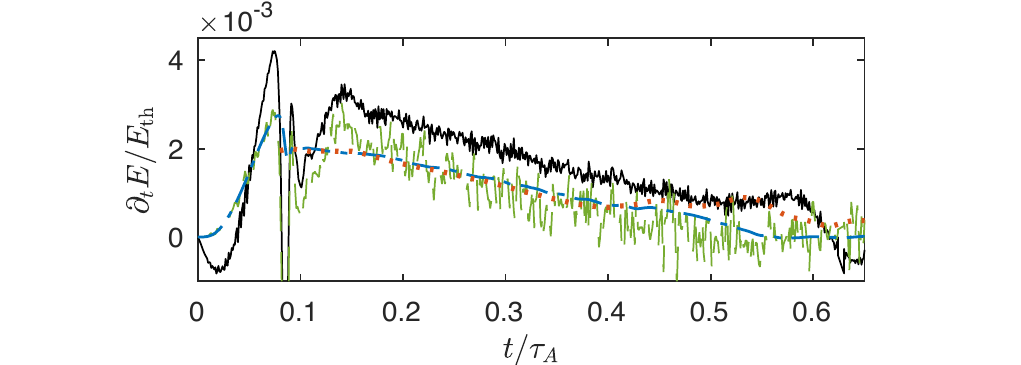}
\caption{Plasma heating due to the standing wave in Fig.~\ref{fig: spacetime Bz standing}. We compare the rate of change of thermal energy $\partial_{t} E_{\mathrm{th}}= \int\!d\bm{x}\,n_{i}\sum_{r} \partial_{t}(\Pi_{rr}/n_{i})/2$ (black line; $\Pi_{rs}$ is the pressure tensor), with mechanical heating -$\int d\bm{x}\,\sum_{rs}\Pi_{rs}\nabla_{r}u_{s}$ (green dashed line),  heating from the large-scale
SA wave $\int\!d\bm{x}\,\Delta\bar{p}\,\hat{b}_{x}\hat{b}_{z}\partial_{x}\bar{u}_{z}$ (blue dot-dashed line; here $\bar{\cdot}$ denotes a filter that smooths fluctuations with $k\rho_{i}\gtrsim 0.25$),
and the approximate viscous heating \cite{Kunz:2010gv} from the SA wave after interruption $\nu_{c}^{-1}\int\!d\bm{x}\,\bar{p}_{\parallel}(\hat{b}_{x}\hat{b}_{z}\partial_{x}\bar{u}_{z})^{2}$ (red dotted line; we use $\nu_{c}/\omega_{A}\approx 10$ as in Fig.~\ref{fig: collisionality and Bz standings}).We normalize by $E_{\mathrm{th}}$ and use units of $\tau_{A}$ (note the small rates, due to 
the high $\beta$).
The initial $\partial_{t}E_{\mathrm{th}}<0$ is due to the creation of $\bm{E}$ fluctuations (because of particle noise).}
\label{fig: plasma heating}
\end{center}
\end{figure}

The subsequent evolution of the oblique firehose modes
controls the large-scale wave dynamics. If these (now stable) fluctuations 
scatter particles sufficiently strongly, $\delta B_{z}$ can  decay with $\Delta p\approx-B^{2}/4\pi$;
if they do not (e.g., if they are resonantly damped \cite{Foote:1979ee,1998ApJ...500..978Q,Melville:2015tt}), $\delta B_{z}$ cannot decrease \cite{Squire:2016ev}.
The firehose
modes' evolution is governed by $\Delta p$ \cite{Rosin:2011er}, which varies in space. Near the wave nodes, where $S=|\nabla \bm{u}|\approx 0$ and $\delta B_{z}\approx 0$, $\Delta p$ is not driven by a large-scale $dB/dt$  and can freely decay \cite{1996JGR...10124457Q,Schekochihin:2008en,doi:10.1063/1.4905230,Melville:2015tt}. 
 Near the wave antinodes, where $S \sim \beta^{\,-1/2}\omega_{A} \approx 6\times 10^{-5}\Omega_{i}$ \cite{Squire:2016ev2}
 and $\delta B_{z}\neq 0$, $\Delta p$ is continuously driven by the decreasing field \cite{2006JGRA..11110101M,Hellinger:2008hd,Kunz:2014kt,Melville:2015tt,2016ApJ...824..123R}.

Surprisingly, it is  small-scale modes at the SA-wave 
nodes---the  \emph{least} firehose-unstable regions (with $4\pi \Delta p/B^{2}\approx -0.7$)---that cause the strongest particle scattering. This is illustrated by the background color in Fig.~\ref{fig: collisionality and Bz standings}, which shows the effective ion collisionality $\nu_{c}/\omega_{A}$ as a function of space,  measured by calculating the time it takes for $\mu$ to change by a factor of $1.2$ for 2048 sample ions \footnote{We then rescale $\nu_{c}$  
so that $\nu_{c}^{-1}$ measures factor-$e$ changes in $\mu$.}. 
The scattering changes from being homogenous and weak at early times, to being stronger and localized 
around the SA-wave nodes at later times.
 This change is caused by the decay of oblique firehose modes into
 $k_{\parallel}\rho_{i}\sim 1,\,k_{\perp}\sim 0$ fluctuations [Fig.~\ref{fig: spacetime Bz standing}(c--d)], which scatter particles efficiently due to their small scale.
 These parallel modes are long lived (they are nonlinearly stabilized against cyclotron damping  \cite{Gary:2003cq,Gary:2004br}), as indicated by their presence after the large-scale SA wave has decayed and $\Delta p \sim 0$ [Fig.~\ref{fig: spacetime Bz standing}(d)].

Because $\omega_{A}\ll\nu_{c}\ll\Omega_{i}$, the plasma dynamics now resemble the Braginskii collisional limit \cite{Braginskii:1965vl} and the SA wave behaves as discussed in  \cite{Squire:2016ev2}.
 We illustrate the similarity in Fig.~\ref{fig: collisionality and Bz standings}(b), which also 
shows $\delta B_{z}$ and $4\pi\Delta p/B^{2}$ for an SA wave governed by the Braginskii model (including heat fluxes; see \cite{Squire:2016ev2}, App.~B).
The ``humped'' shape occurs because the perturbation splits into regions where 
$4\pi\Delta p\approx-B^{2}$ and $d\delta B_{z}/dt<0$ (around the antinodes), and regions where $4\pi\Delta p>-B^{2}$ and $\delta B_{z}=0$ (these spread from the  nodes).
The wave decay rate is determined by $\nu_{c}$ which is sufficiently large 
[$\nu_{c}/\omega_{A}\sim\beta(\delta B_{\perp}/B_{0})^{2}$] that the wave decays 
within $t/\tau_{A}<1$. 
We note parenthetically that the wave decay generates a $\delta B_{y}$ perturbation (see Fig.~\ref{fig: spacetime Bz standing}), although its origin is currently unclear.

\begin{figure}
\begin{center}
\includegraphics[width=1.0\columnwidth]{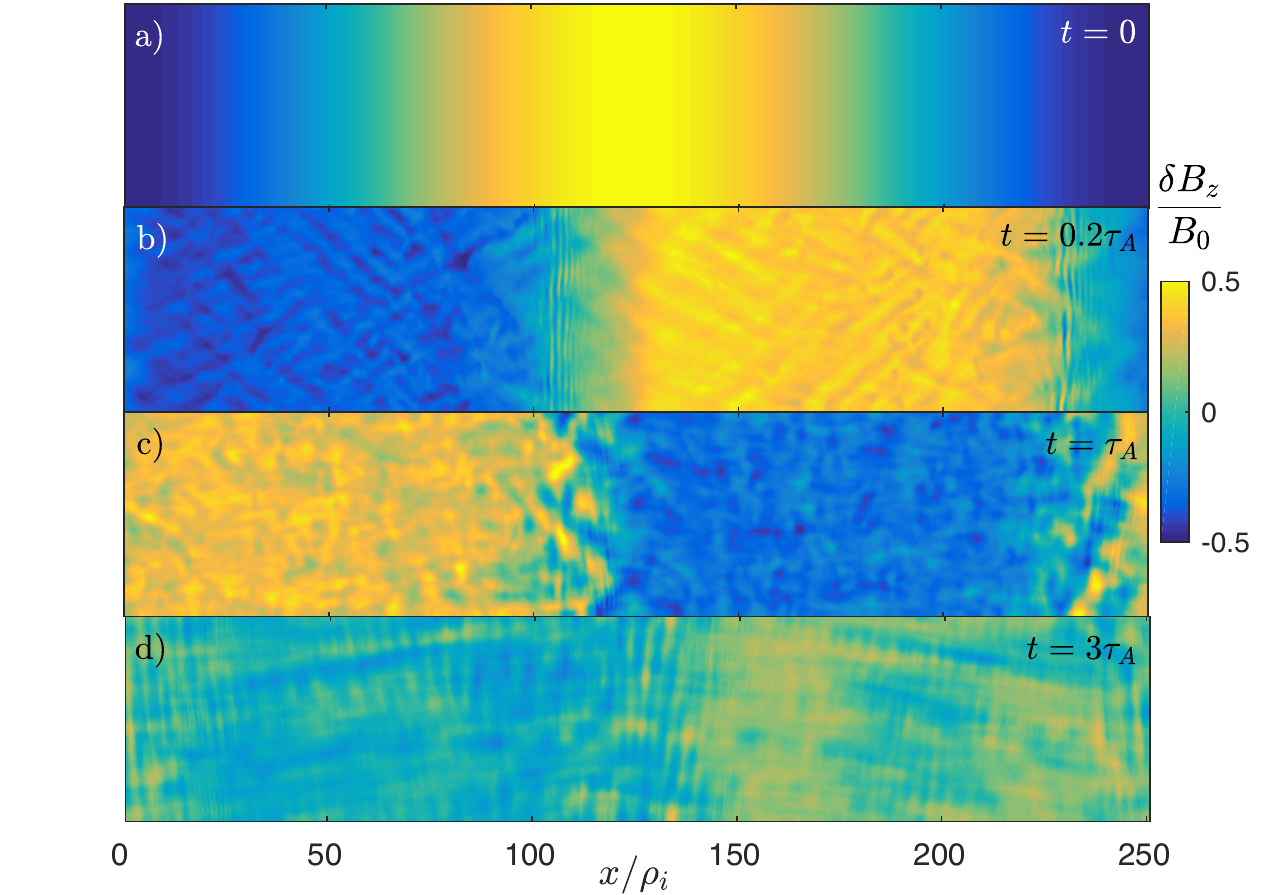}
\caption{Out-of-plane magnetic perturbation $\delta B_{z}/B_{0}$ for a SA traveling wave with $\lambda_{A}=250\rho_{i}$ at  $t=0$ (a), $t=0.2\tau_A$ (b), $t=\tau_A$ (c), and $t=3\tau_A$ (d).
}
\label{fig: traveling spacetime Bz}
\end{center}
\end{figure}

As the large-scale SA wave decays, it heats 
the plasma. This process does not 
involve a turbulent cascade, but rather  the 
direct transfer of large-scale mechanical energy into thermal energy.
This heating is essentially viscous dissipation, with particle scattering
from microscale fluctuations controlling the effective viscosity and making the process irreversible.
In Fig.~\ref{fig: plasma heating}, we compare the measured $\partial_{t} E_{\mathrm{th}}$ with heating due to the SA wave decay. Although the agreement is not perfect due to spurious   grid heating \cite{1991ppcs.book.....B} (tests at  
$\lambda_{A}/\rho_{i}=250$ show that this improves with ppc or reduced $\lambda_{A}/\rho_{i}$), the  stages of wave decay are evident; e.g., the  drop 
in $\partial_{t} E_{\mathrm{th}}$ as firehose fluctuations grow at $t/\tau_{A}\approx 0.075$, followed by heating as the large-scale $\delta B_{\perp}$ decays. Fig.~\ref{fig: plasma heating} also shows that the overall energetics are well captured by considering only the large-scale dynamics, or 
by using the same effective collisionality as in Fig.~\ref{fig: collisionality and Bz standings}(b).
This supports  closure models that approximate the effects of
microinstabilities on large-scale dynamics without having to resolve the microscales.

\begin{figure}
\begin{center}
\includegraphics[width=1.0\columnwidth]{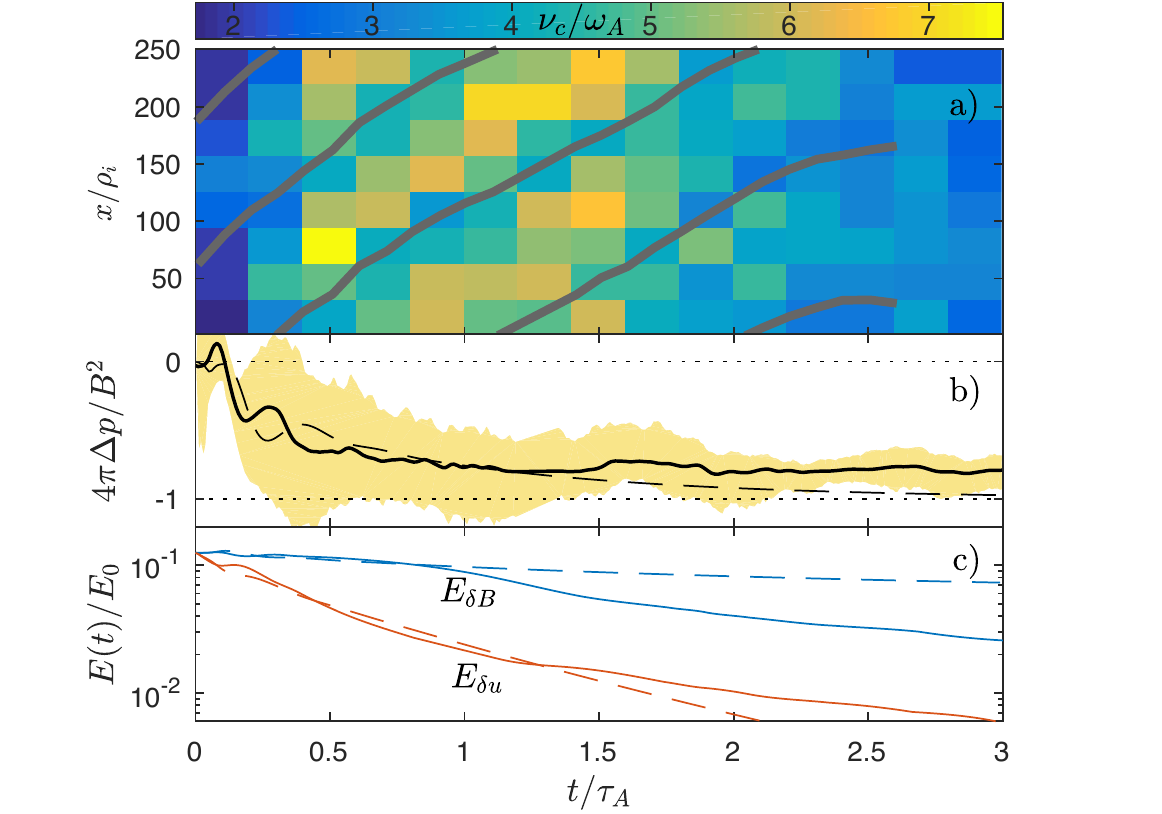}
\caption{(a) Scattering rate $\nu_{c}/\omega_{A}$ of the traveling wave in Fig.~\ref{fig: traveling spacetime Bz} as a function of $x$ and $t$.
The grey lines follow the wave fronts (this is 
close to where $ \Delta p$ is most negative).
 (b) Time evolution of $\langle 4 \pi \Delta p/ B^{2}\rangle$. The shaded region indicates the range of  $4 \pi \Delta p/ B^{2}$ seen across the wave profile, to illustrate when the wave can excite firehose modes. (c) Energy of the magnetic perturbation $E_{\delta B} =\int d\bm{x}\,\delta B_{z}^{2}/8\pi$ (blue) and kinetic energy $E_{\delta u}=\int d\bm{x}\,\rho\delta u_{z}^{2}/2$ (red), normalized by $E_{0}=\int d\bm{x}\,B_{0}^{2}/8\pi$. In (b) and (c), we plot 
the results from an equivalent Landau-fluid simulation \cite{Squire:2016ev2} (dashed lines), for comparison.}
\label{fig: traveling 1D profiles}
\end{center}
\end{figure}

\emph{Shear-Alfv\'en traveling wave.---}Figures~\ref{fig: traveling spacetime Bz} and \ref{fig: traveling 1D profiles} illustrate the dynamics 
of the nonlinear SA traveling wave with $\lambda_{A} = 250\rho_{i}$. The initial evolution differs  
from standing waves because $\langle d B/dt \rangle=0$ for an unperturbed 
traveling wave, implying that global (spatially constant) pressure anisotropy is  created only as the wave 
decays \cite{Squire:2016ev}. The evolution broadly follows the expectations of \cite{Squire:2016ev,Squire:2016ev2}, proceeding
in 4 stages: (i) the spatially dependent $dB/dt$ creates an anisotropy $\Delta p (x) \sim \beta^{1/2 } \delta B_{z}^{2} \sin(2 k_{\parallel}x ) $; (ii) this $\Delta p$ damps the wave  \cite{Squire:2016ev2,Lee:1973ky,Kulsrud:1978cr} causing $\langle B \rangle$ to decrease and thus $\langle \Delta p\rangle<0$; (iii) the wave consequently slows down, 
with $\delta u_{\perp}$
decaying faster than $\delta B_{\perp}$; and (iv) the wave excites oblique firehose 
modes, which subsequently scatter particles and cause the large-scale $\delta B_{\perp}$ to decay in a similar manner to the standing wave.

These stages are seen clearly in Figs.~\ref{fig: traveling spacetime Bz} and \ref{fig: traveling 1D profiles}.
In particular, note the global $\langle \Delta p \rangle<0$ that quickly develops [Fig.~\ref{fig: traveling 1D profiles}(b)] and the fast decay of $\delta u_{\perp}$ at early times [Fig.~\ref{fig: traveling 1D profiles}(c)].  By $t/\tau_{A}\approx 0.3$, once $\langle\Delta p\rangle$ has decreased sufficiently, the wave starts exciting oblique firehose modes.
Unlike for standing waves, this occurs only in  localized regions around the wavefronts (i.e., near where $\delta B_{z}=0$),
because $|dB/dt |$ (and thus $|\Delta p|$) is largest in these regions  [see, e.g., Fig.~\ref{fig: traveling spacetime Bz}(c) at $x/\rho_{i}\approx 110$ and the shading in Fig.~\ref{fig: traveling 1D profiles}(b)].
Subsequently, Fig.~\ref{fig: traveling 1D profiles}(a)  shows that the particle scattering 
is strongest behind the wavefronts. 
We interpret this as being due to the transition of oblique firehose modes into long-lived  $k_{\parallel}\rho_{i}\sim 1 $ fluctuations [see Fig.~\ref{fig: traveling spacetime Bz}(d)], like in the standing wave. Because firehose modes are excited only briefly around the wavefronts, the scattering rate 
$ \nu_{c}/\omega_{A} $ is lower than for the standing wave. Thus the final decay of the traveling wave's $\delta B_{z}$ is slower than the standing wave's,  although it is qualitatively similar. 
At earlier times, the large-scale SA wave evolution matches well the predictions from a 1-D LF model at $\beta_{i}=100$ \cite{Squire:2016ev2} (dashed lines in Fig.~\ref{fig: traveling 1D profiles}).


\emph{Discussion.---}We have presented hybrid-kinetic simulations of large-amplitude SA waves in a collisionless 
plasma. 
Our results  demonstrate clearly the exceptional influence of microinstabilities on the large-scale ($\lambda_{A} \gg \rho_{i}$) dynamics  of high-$\beta$ collisionless plasmas, illustrating 
how the evolution of self-excited oblique firehose modes \emph{controls} the plasma's fluid properties.   The simulations also verify, using a  realistic model with  kinetic ions, that linearly polarized shear-Alfv\'enic perturbations do not exist in their linear wave form above the amplitude limit
$\delta B_{\perp}/B_{0}\sim \beta^{\,-1/2}$ \cite{Squire:2016ev}.
The SA wave dynamics depend strongly on how oblique firehose modes evolve as the plasma becomes stable ($\Delta p\gtrsim-B^{2}/4\pi$). 
We find that firehose fluctuations become parallel ($k_{\perp}=0$) 
 and move to smaller scales ($k_{\parallel}\rho_{i}\sim 1$), surviving nonlinearly throughout the large-scale $\delta B_{\perp}$ decay and  scattering particles at a high rate. These long-lived $k_{\parallel}\rho_{i}\sim 1$ modes cause SA standing-wave dynamics in 
a collisionless plasma to  resemble  those in a collisional (Braginskii) one \cite{Squire:2016ev2}.
The initial evolution of the traveling wave is effectively collisionless and matches analytic predictions \cite{Squire:2016ev2}; however,
after generating a global negative anisotropy and exciting firehose modes, its  final decay resembles the standing 
wave.
For both standing and traveling waves, the simulations
provide an interesting example of  direct transfer of energy from the largest scales to thermal energy and microscale fluctuations, without  a turbulent cascade.

 
 Our simulations cannot fully address what occurs at yet higher $\lambda_{A}/\rho_{i}$. This will depend on how oblique firehose modes decay and scatter particles, physics that is currently poorly understood.   That said, it is clear that SA wave interruption provides a robust mechanism for  dissipating energy directly from large-scale perturbations into heat and microinstabilities.   
 Our results suggest that numerical models of weakly collisional high-$\beta$ plasmas would be better off damping large-amplitude SA waves, rather than letting them freely propagate.   One concrete way to achieve  this aim might be a LF model with  pressure-anisotropy limiters \cite{Sharma:2006dh} that enhance 
 the collisionality to a rate that is determined by the large-scale Alfv\'en frequency.
 More work on developing and validating subgrid models of this kind is underway. 
 
Given the strong deviations from MHD predictions, SA wave interruption could significantly 
impact the turbulent dynamics of weakly collisional plasmas in a variety of astrophysical environments \cite{Kunz:2010gv}. 
 Some  effects 
 have already been observed in the $\beta \sim 1$ solar wind \cite{BalePrep}. Other astrophysical plasmas---for instance the ICM, with $\beta\sim 100$ \cite{2006A&A...453..447E,Rosin:2011er}---are likely to be more strongly affected by interruption, and work is underway to assess its impact on turbulence under such conditions.

\begin{acknowledgments}
We thank S.~Balbus, S.~D.~Bale, C.~H.~K~Chen, S.~Cowley, B.~Dorland, G.~Hammett,    K.~Klein,  F.~Rincon, L.~Sironi, and M.~Strumik for useful and enlightening discussions. JS, AAS, and MWK would like to thank the Wolfgang Pauli Institute in Vienna for its hospitality on several occasions. JS was funded in part by the Gordon and Betty Moore Foundation
through Grant GBMF5076 to Lars Bildsten, Eliot Quataert and E. Sterl
Phinney. EQ was supported by Simons Investigator awards from the Simons Foundation and  NSF grant AST 13-33612. AAS was supported in part by grants from  UK STFC and EPSRC. MWK was supported in part by NASA grant NNX16AK09G
and US DOE Contract DE-AC02-09-CH11466.
This work used the Extreme Science and Engineering Discovery Environment (XSEDE), which is supported by National Science Foundation grant number ACI-1548562. Numerical calculations were carried out on the Comet system at the San Diego supercomputing center, through allocation TG-AST160068.
\end{acknowledgments}

\bibliographystyle{apsrev4-1}
\bibliography{fullbib,bib_extrapapers}

\end{document}